\documentstyle[aps]{revtex}                

\draft 		

\input{epsf}

\begin{document}


\def\issue(#1,#2,#3){{\bf #1}, #2 (#3)} 

\def\opcit(#1){ {\em op. cit.}, #1}

\def\ARNPS(#1,#2,#3){Ann.\ Rev.\ Nucl.\ Part.\ Sci.\ \issue(#1,#2,#3)}
\def\CPC(#1,#2,#3){Comp.\ Phys.\ Comm.\ \issue(#1,#2,#3)}
\def\EPJC(#1,#2,#3){Eur.\ Phys.\ J.\ C\ \issue(#1,#2,#3)}
\def\IEEETNS(#1,#2,#3){IEEE Trans.\ Nucl.\ Sci.\ \issue(#1,#2,#3)}
\def\NP(#1,#2,#3){Nucl.\ Phys.\ \issue(#1,#2,#3)}
\def\NIM(#1,#2,#3){ Nucl.\ Inst.\ and Meth.\ \issue(#1,#2,#3)}
\def\PL(#1,#2,#3){Phys.\ Lett.\ \issue(#1,#2,#3)}
\def\PRD(#1,#2,#3){Phys.\ Rev.\ D \issue(#1,#2,#3)}
\def\PRL(#1,#2,#3){Phys.\ Rev.\ Lett.\ \issue(#1,#2,#3)}
\def\SJNP(#1,#2,#3){Sov.\ J. Nucl.\ Phys.\ \issue(#1,#2,#3)}
\def\ZPC(#1,#2,#3){Zeit.\ Phys.\ C \issue(#1,#2,#3)}

\title{ 
\begin{flushright}{\normalsize FERMILAB-Pub-99/036-E}\end{flushright}
Measurements of Lifetimes and a Limit on the Lifetime 
	Difference in the Neutral $D$-Meson System
}
\author{
    E.~M.~Aitala,$^9$
       S.~Amato,$^1$
    J.~C.~Anjos,$^1$
    J.~A.~Appel,$^5$
       D.~Ashery,$^{14}$
       S.~Banerjee,$^5$
       I.~Bediaga,$^1$
       G.~Blaylock,$^8$
    S.~B.~Bracker,$^{15}$
    P.~R.~Burchat,$^{13}$
    R.~A.~Burnstein,$^6$
       T.~Carter,$^5$
    H.~S.~Carvalho,$^{1}$
    N.~J.~Copty,$^{12}$
    L.~M.~Cremaldi,$^9$
       C.~Darling,$^{18}$
       K.~Denisenko,$^5$
       A.~Fernandez,$^{11}$
    G.~F.~Fox,$^{12}$,
       P.~Gagnon,$^2$
       C.~Gobel,$^1$
       K.~Gounder,$^9$
    A.~M.~Halling,$^5$
       G.~Herrera,$^4$
       G.~Hurvits,$^{14}$
       C.~James,$^5$
    P.~A.~Kasper,$^6$
       S.~Kwan,$^5$
    D.~C.~Langs,$^{12}$
       J.~Leslie,$^2$
       B.~Lundberg,$^5$
       S.~MayTal-Beck,$^{14}$
       B.~Meadows,$^3$
 J.~R.~T.~de~Mello~Neto,$^1$
       D.~Mihalcea,$^7$
    R.~H.~Milburn,$^{16}$
    J.~M.~de~Miranda,$^1$
       A.~Napier,$^{16}$
       A.~Nguyen,$^7$
    A.~B.~d'Oliveira,$^{3,11}$
       K.~O'Shaughnessy,$^2$
    K.~C.~Peng,$^6$
    L.~P.~Perera,$^3$
    M.~V.~Purohit,$^{12}$
       B.~Quinn,$^9$
       S.~Radeztsky,$^{17}$
       A.~Rafatian,$^9$
    N.~W.~Reay,$^7$
    J.~J.~Reidy,$^9$
    A.~C.~dos Reis,$^1$
    H.~A.~Rubin,$^6$
    D.~A.~Sanders,$^9$
 A.~K.~S.~Santha,$^3$
 A.~F.~S.~Santoro,$^1$
       A.~J.~Schwartz,$^{3}$
       M.~Sheaff,$^{17}$
    R.~A.~Sidwell,$^7$
    A.~J.~Slaughter,$^{18}$
    M.~D.~Sokoloff,$^3$
       J.~Solano,$^1$
    N.~R.~Stanton,$^7$
    R.~J.~Stefanski,$^5$  
       K.~Stenson,$^{17}$ 
    D.~J.~Summers,$^9$
       S.~Takach,$^{18}$
       K.~Thorne,$^5$
    A.~K.~Tripathi,$^{7}$
       S.~Watanabe,$^{17}$
 R.~Weiss-Babai,$^{14}$
       J.~Wiener,$^{10}$
       N.~Witchey,$^7$
       E.~Wolin,$^{18}$
    S.~M.~Yang,$^7$
       D.~Yi,$^9$
       S.~Yoshida,$^7$
       R.~Zaliznyak,$^{13}$ and
       C.~Zhang$^7$ \\
\begin{center} (Fermilab E791 Collaboration) \end{center}
}

\address{
$^1$ Centro Brasileiro de Pesquisas F\'\i sicas, Rio de Janeiro, Brazil\\
$^2$ University of California, Santa Cruz, California 95064\\
$^3$ University of Cincinnati, Cincinnati, Ohio 45221\\
$^4$ CINVESTAV, Mexico\\
$^5$ Fermilab, Batavia, Illinois 60510\\
$^6$ Illinois Institute of Technology, Chicago, Illinois 60616\\
$^7$ Kansas State University, Manhattan, Kansas 66506\\
$^8$ University of Massachusetts, Amherst, Massachusetts 01003\\
$^9$ University of Mississippi, University, Mississippi 38677\\
$^{10}$ Princeton University, Princeton, New Jersey 08544\\
$^{11}$ Universidad Autonoma de Puebla, Mexico\\
$^{12}$ University of South Carolina, Columbia, South Carolina 29208\\
$^{13}$ Stanford University, Stanford, California 94305\\
$^{14}$ Tel Aviv University, Tel Aviv, Israel\\
$^{15}$ Box 1290, Enderby, British Columbia, V0E 1V0, Canada\\
$^{16}$ Tufts University, Medford, Massachusetts 02155\\
$^{17}$ University of Wisconsin, Madison, Wisconsin 53706\\
$^{18}$ Yale University, New Haven, Connecticut 06511
}

\maketitle

\begin{abstract}
Using the large hadroproduced charm sample collected in experiment E791
at Fermilab, we report the first directly measured constraint on the
decay-width difference $\Delta \Gamma$ for the mass eigenstates of the
$D^0$-$\overline D\,^0$ system.  We obtain our result from lifetime
measurements of the decays $D^0\rightarrow K^-\pi^+$\ and $D^0
\rightarrow K^-K^+$, under the assumption of {\em CP} invariance, which
implies that the {\em CP} eigenstates and the mass eigenstates are the
same. The lifetime of $D^0 \rightarrow K^-K^+$ (the {\em CP}-even final
state) is $\tau_{K\!K} = 0.410 \pm 0.011 \pm 0.006$~ps, and the
lifetime of $D^0\rightarrow K^-\pi^+$\ (an equal mixture of {\em
CP}-odd and {\em CP}-even final states) is $\tau_{K\!\pi} = 0.413 \pm
0.003 \pm 0.004$~ps.  The decay-width difference is $\Delta \Gamma =
2(\Gamma_{K\!K} - \Gamma_{K\!\pi}) = 0.04 \pm 0.14 \pm 0.05$
ps$^{-1}$.  We relate these measurements to measurements of mixing in
the neutral $D$-meson system.
\end{abstract}


The two mass eigenstates of the $D^0$-$\overline D\,^0$ system may have
different lifetimes. In the Standard Model the dominant contribution to
such a difference arises from box-diagram amplitudes involving the
exchange of two $W$ bosons.  Suppression factors due to the GIM
mechanism and heavy quark symmetries in QCD lead to the conclusion that
these lifetime and mass differences are too small to be measured
\cite{SMboxcharm}. Additional Standard Model contributions from penguin
diagrams \cite{petrov} and long-distance effects \cite{SMlongdist} are
also expected to be negligible.  Thus, a measurable difference in the
lifetimes \cite{liu} or masses would constitute evidence for new
physics.  In this letter, we report the first direct search for a
lifetime difference in the neutral $D$-meson system.

Because lifetime and mass differences lead to particle-antiparticle
mixing, previous experiments have looked for mixing as evidence of such
differences.  The rate of mixing is usually characterized by the
parameter $r_{\rm mix}$, which is defined as the ratio of mixed to
unmixed decays
\begin{equation}
\label{rmix} 
r_{\rm mix} \equiv 
{\Gamma(D^0\rightarrow\overline D\,^0\rightarrow \overline{f}) \over 
\Gamma(D\,^0\rightarrow f)}\,. 
\end{equation} 
This ratio can have contributions from both the mass difference $\Delta
m$ and the decay-rate difference $\Delta \Gamma$ between the two
neutral $D$ mass eigenstates according to 
\begin{equation}
\label{r_mix_mg} 
r_{\rm mix} = {(\Delta m)^2\over 2\Gamma^2} +  
{(\Delta\Gamma)^2\over 8\Gamma^2}\,, 
\end{equation} 
where $\Gamma$ is the average of the decay rates of the two neutral $D$
mass eigenstates.  

To date, particle-antiparticle mixing has only been observed in the
strange- and bottom-quark sectors.  Previous searches for
$D^0$-$\overline D\,^0$ mixing have set a limit $r_{\rm mix} < 0.50\%$
at 90\% confidence level (CL) \cite{prevmixing}. Inserting this value
into Eq.\ \ref{r_mix_mg} and assuming $\Delta m = 0$, one obtains the
limit $\left|\Delta \Gamma\right| < 0.48$ ps$^{-1}$ at 90\% CL
\cite{pdg}.  The measurement of $\Delta \Gamma$ reported here is more
sensitive than these previous measurements. 

If {\em CP} conservation holds in the $D^0$-meson system, the even and
odd {\em CP} eigenstates $D^0_1$ and $D^0_2$ will be the mass
eigenstates; {\em i.e.}, they will have definite masses $m_1$ and
$m_2$, and decay widths $\Gamma_1$ and $\Gamma_2$. Since the
Cabibbo-suppressed final state $K^-K^+$ is {\em CP}-even, it results
from the decay\footnote{We imply charge-conjugate decay modes
throughout this letter.} of $D^0_1$ and will have an exponential
decay-rate distribution
\begin{equation}
\label{gkk}
\Gamma_{K\!K}(t) = A_{K\!K}e^{-\Gamma_1 t}
\end{equation}
where here and in Eqs.~\ref{gkpi} and \ref{gkpiexp} below, $A$
represents all time-independent factors. In contrast, the
Cabibbo-favored final state $K^-\pi^+$ receives roughly equal
contributions from $D^0_1$ and $D^0_2$\@. Under the assumption of {\em
CP} conservation and ignoring doubly-Cabibbo-suppressed terms, the
decay-rate distribution of a combined sample of $D^0 \rightarrow
K^-\pi^+$\ and $D^0 \rightarrow K^+\pi^-$ (from mixing), and separately
a combined sample of $\overline D\,^0 \rightarrow K^+\pi^-$\ and
$\overline D\,^0 \rightarrow K^-\pi^+$, can be described by
\cite{dunietz,palmerwu}\footnote{Reference \cite{palmerwu} omits the
factor of 2 in the $\cosh$ term.}
\begin{equation}
\label{gkpi}
\Gamma_{K\!\pi}(t) = A_{K\!\pi}e^{-\Gamma t}\cosh\left(\frac{\Delta 
\Gamma}{2}t\right)\!,
\end{equation}
where
\begin{equation}
\label{dgave}
\Gamma = (\Gamma_1 + \Gamma_2)/2 \mbox{\ \ and\ \ } \Delta \Gamma 
= \Gamma_1 - \Gamma_2\,,
\end{equation}
and so
\begin{equation}  
\label{dgdiff}  
{\Delta\Gamma\over 2} = \Gamma_1 -
\Gamma = \Gamma_{K\!K} - \Gamma_{K\!\pi} =  {1\over\tau_{K\!K}} -
{1\over\tau_{K\!\pi}}\,.  
\end{equation}
Given the current limits on $\Delta \Gamma$, Eq.~\ref{gkpi} is
well-approximated by a pure exponential distribution
\begin{equation}
\label{gkpiexp}
\Gamma_{K\!\pi}(t) = A_{K\!\pi}e^{-\Gamma t}
\end{equation}
over the range of sensitivity of our experiment.
  
If the mass states are {\em not} {\em CP} eigenstates, {\it i.e.}, {\em
CP} violation is present, then interpreting the measured lifetime
difference in terms of $\Gamma_1$ and $\Gamma_2$ is less
straightforward.  However, even in the absence of strict {\em CP}
conservation, pure exponential distributions are still good
approximations for both the $KK$ and $K\pi$ distributions in our data
sample. In this case, a measured difference at our level of sensitivity
would constitute evidence for physics beyond the Standard Model,
although it might not be easily related to the lifetime difference of
the mass eigenstates. For convenience of interpretation, we hereafter
assume {\em CP} invariance in the $D^0$-meson system, though the
validity of that assumption has previously been questioned in the
literature \cite{CPdiscussion}.

The results reported here are based on data accumulated by experiment
E791 in a 500-GeV/$c$ $\pi^-$ beam during the 1991/92 Fermilab
fixed-target run. E791 was the fourth in a series of charm experiments
performed in the Fermilab Tagged Photon Laboratory. The E791
spectrometer \cite{exp791} was an open-geometry detector with 23 planes
of silicon microstrip detectors (6 upstream and 17 downstream of the
target), 35 drift chamber planes, 10 proportional wire chambers (8
upstream and 2 downstream of the target), two magnets for momentum
analysis, two large multicell threshold \v{C}erenkov counters for
charged particle identification, electromagnetic and hadronic
calorimeters for electron/hadron separation as well as for online
triggering, and a fast data acquisition system that  collected data at
a rate of up to 30 Mbyte/s with a 50~$\mu$s/event deadtime. The target
consisted of a 0.52-mm platinum foil followed by four 1.6-mm diamond
foils. Each target center was separated from the next by about 1.5 cm,
allowing observation of charm-particle decays in air with minimal
background from secondary interactions in the targets. The very loose
transverse-energy trigger was based on the energy deposited in the
calorimeters and was highly efficient for charm events. Over $2 \times
10^{10}$ events were recorded during a six-month period.

The measurement of $\Delta \Gamma$ is based on the observation of $D^0
\rightarrow K^-K^+$ and $K^-\pi^+$ decay modes in the same detector
with similar detector and topological biases and uncertainties, leading
to a precise comparison of the lifetimes for the two modes. This is a
companion analysis to that of Ref.\ \cite{kkpipi}, in which we set
limits on {\em CP} violation and measured the branching ratios for $D^0
\rightarrow K^-K^+$ and $D^0 \rightarrow \pi^-\pi^+$ relative to that
for  $D^0 \rightarrow K^-\pi^+$.

Selection requirements for inclusion in the $D^0 \rightarrow K^-K^+$
and $D^0 \rightarrow K^-\pi^+$ samples included a minimum significance
of separation of the candidate decay vertex from the production vertex
in the beam direction ($ > 8 \sigma$ where $\sigma$ is the uncertainty
in the separation of the two vertices).  Candidate decay tracks were
required to be inconsistent with coming from the $D^0$ production
vertex. The net momentum of the $D^0$ candidate transverse to the line
connecting the production and decay vertices was required to be less
than 0.35 GeV/$c$, and the sum of the $p_t^2$ of the decay tracks was
required to be greater than 0.25 (${\rm GeV}/c)^2$, with $p_t$ measured
relative to the direction of the candidate $D^0$. Finally, to remove
secondary interactions, the decay vertex had to be located at least
4$\sigma$ outside the target foils, where $\sigma$ is the uncertainty
in the position of the decay vertex in the beam direction. 

\begin{figure}[htb]
\centerline {\epsfxsize=4.75in \epsffile{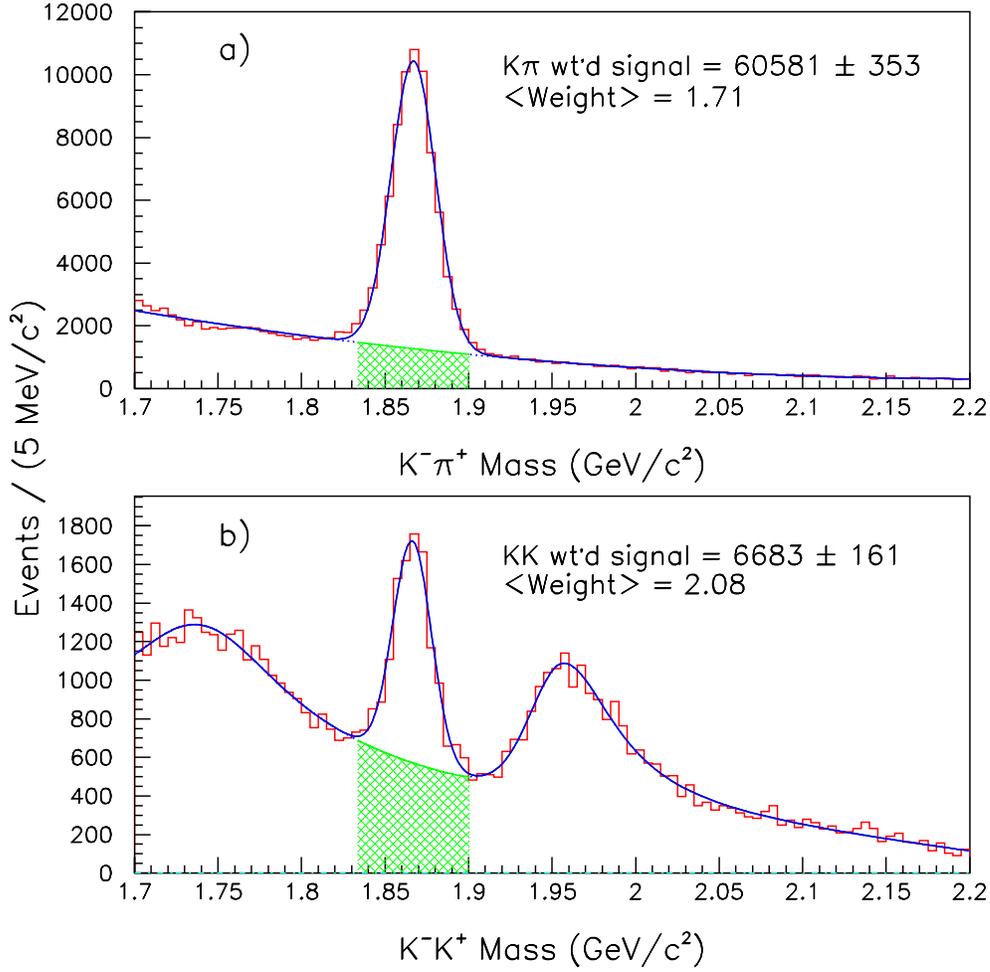} }
\caption{
Invariant mass distribution for candidates weighted by particle 
identification efficiency (see text).
(a) $D^0 \rightarrow K^-\pi^+$ decays. The solid line corresponds to
a fit to a Gaussian signal distribution plus a third-order polynomial
for combinatoric background.    
(b) $D^0 \rightarrow K^-K^+$ decays. The solid line corresponds
to a fit to a Gaussian signal distribution plus an asymmetric
Breit-Wigner ($D^0 \rightarrow K^-\pi^+$ reflection), a symmetric
Breit-Wigner ($D^0 \rightarrow K^-\pi^+\pi^0$ reflection), and a linear
combinatoric background.  The cross-hatched areas are the estimated
backgrounds beneath our signal region of width $\pm 2.5$ Gaussian
standard deviations about the Gaussian mean.
}
\label{fg:tausignal}
\end{figure}

We used particle identification from the \v{C}erenkov counters to
improve the statistical significance of the $D^0 \rightarrow K^-K^+$
sample.  To minimize systematic effects in calculating $\Delta \Gamma$,
similar particle identification criteria were applied to the $K^-\pi^+$
sample as well.  We obtained the necessary particle-identification
efficiencies from a study of the $D^0\rightarrow K^-\pi^+$\ sample. We
corrected for the \v{C}erenkov counter efficiency by a weighting
procedure whereby each two-track candidate decay was weighted by the
inverse of the product of the two particle-identification efficiencies
calculated for their individual $p$, $p_t$, and particle type.  Figure
\ref{fg:tausignal} shows the weighted invariant mass plots for our
final data sample. The numbers of unweighted signal events in the
$K^-K^+$ and $K^-\pi^+$ samples are approximately 3200 and 35,400,
respectively.

To account for the effects of our selection criteria, we define a
reduced decay length for each $D^0$ candidate as the distance traveled
by the candidate beyond that required to survive our selection
criteria, and a reduced proper time corresponding to that reduced decay
length. To the extent that the $K^-K^+$ final state is a mass
eigenstate of the neutral $D$-meson system, its proper-decay-time
distribution is purely exponential. In this case, its
reduced-proper-decay-time distribution is also purely exponential with
the same characteristic decay time.  In contrast, the $K\pi$ final
state is a mixture of mass eigenstates with its proper-decay-time
distribution given by Eq.~\ref{gkpi}\@.  However, since the difference
in the decay rates of the eigenstates is small (as known from limits on
$r_{\rm mix}$), the difference in the amount of $D^0_1$ and $D^0_2$
contributing to the $D^0 \rightarrow K^-\pi^+$ decay after our
proper-decay-length cut is negligible.  Also, effects of fitting to a
single exponential of reduced proper time are smaller by an order of
magnitude than the systematic errors described below.  We use reduced
proper lifetime because it minimizes acceptance corrections and
associated systematic errors. We measure the lifetimes over the same
range of reduced proper time for both decay modes. 

For each bin of reduced proper time, candidates are taken from the
total sample shown in Figure \ref{fg:tausignal}, where the signal
region is defined as within $\pm 2.5$ Gaussian standard deviations of
the Gaussian mean, and the background is taken from sidebands, but at
the level of the cross-hatched area.  We use the same $D^0$ mass and
Gaussian width in each of the 16 bins of reduced proper time, as
determined from Figure \ref{fg:tausignal} separately for $D^0
\rightarrow K^-K^+\ (1866.7\ {\rm and}\ 11.6\ {\rm MeV/}c^2)$ and
$D^0\rightarrow K^-\pi^+\ (1867.0\ {\rm and}\ 13.5\ {\rm MeV/}c^2)$.

\begin{figure}[htb]
\centerline {\epsfxsize=6.0in \epsffile{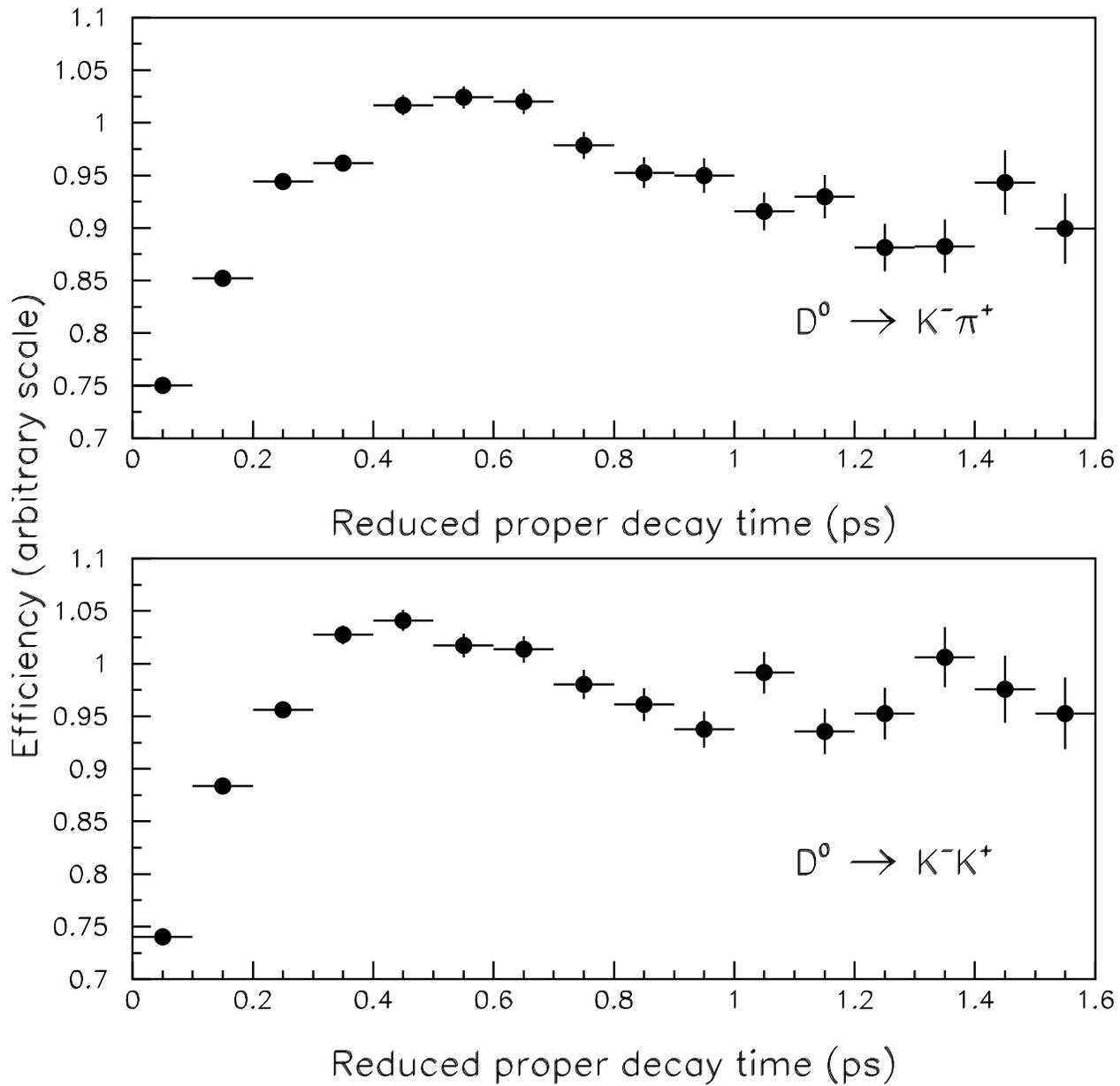} }
\vspace*{-.4in}
\caption{
The $K^-\pi^+$ (top) and $K^-K^+$ (bottom) efficiencies as functions of
the reduced proper decay time. Note that the vertical scales have
suppressed zeros.
} 
\label{fg:accept} 
\end{figure}

\begin{figure}[htb]
\centerline {\epsfxsize=6.0in \epsffile{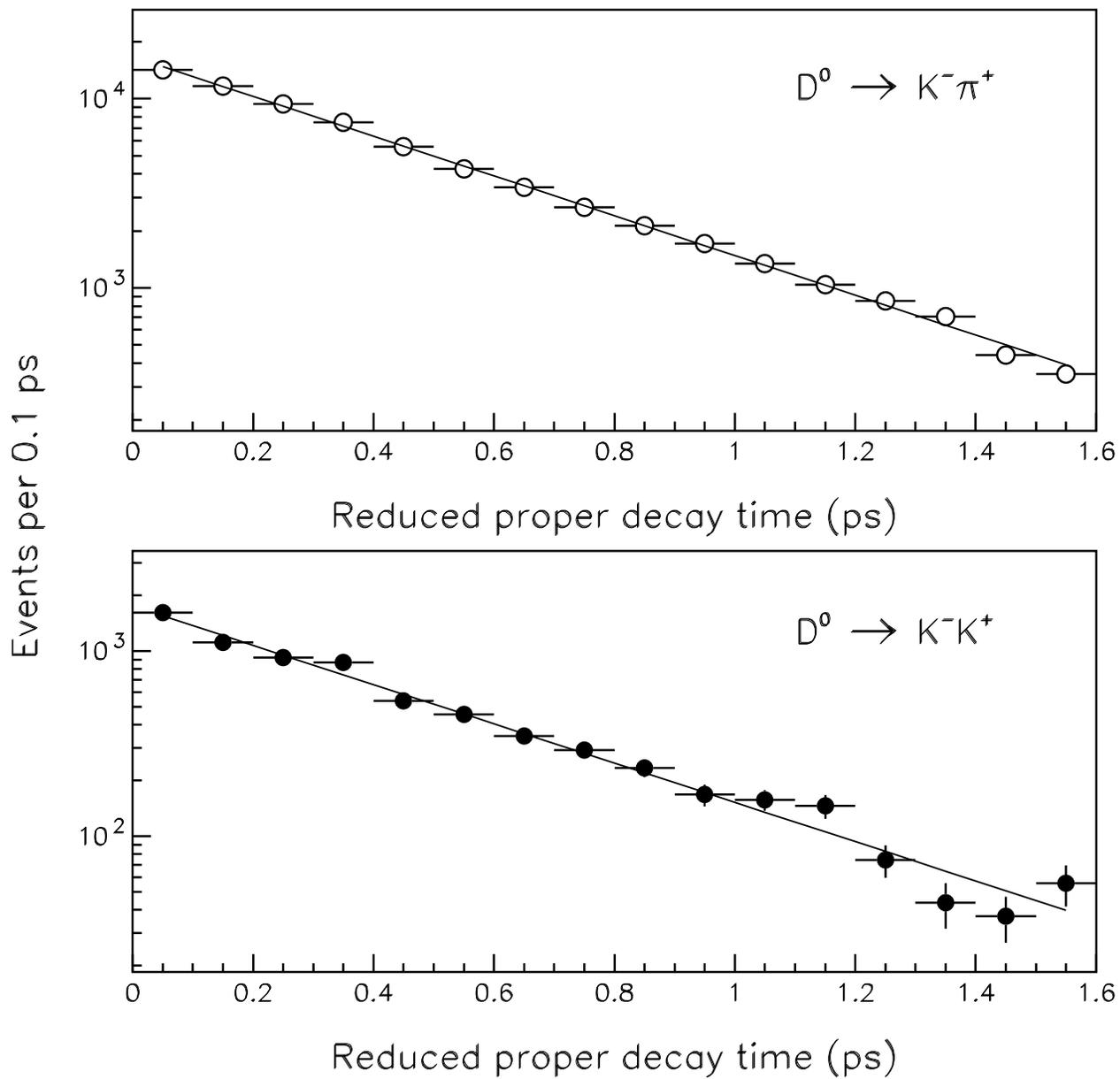} }
\vspace*{-.4in}
\caption[]{
Distributions and the exponential fits for the number of $D^0 \rightarrow
K^-\pi^+$ (top) and $K^-K^+$ (bottom) decays as a function of reduced
proper decay time, after particle identification weighting and
acceptance corrections. 
}
\label{fg:taufits}
\end{figure}

We correct for small acceptance differences between the $K^-K^+$ and
$K^-\pi^+$ final states by using a Pythia-based Monte Carlo model
\cite{pythia} that incorporates a full detector simulation. The
simulated events are subjected to exactly the same  selection criteria
and binning as are the data. Figure \ref{fg:accept} shows the
efficiencies from the Monte Carlo study as functions of the reduced
proper decay time for $K^K-^+$ and $K^-\pi^+$ separately. Figure\
\ref{fg:taufits} presents the exponential fits to the measured
reduced-proper-decay-time distributions after particle-identification
weighting and acceptance corrections.  We measure decay widths
$\Gamma_{K\!K}= 2.441 \pm 0.068$ ps$^{-1}$ and $\Gamma_{K\!\pi} = 
2.420 \pm 0.019$ ps$^{-1}$; or lifetimes $\tau_{K\!K}= 0.410 \pm 0.011$
ps and $\tau_{K\!\pi} = 0.413 \pm 0.003$ ps, where the quoted errors
are statistical only.  Using Eq.~\ref{dgdiff}, we calculate  $\Delta
\Gamma = 0.04  \pm 0.14$  ps$^{-1}$. We have also performed a joint fit
to the distributions using Eqs.~\ref{gkk} and \ref{gkpi} which yields
the same result.

Systematic errors are studied separately for the two lifetimes and for
the lifetime difference. These studies include the effects of varying
the event selection criteria, the $D^0$ production model used in the
Monte Carlo, the particle-identification weighting procedure, the
effect of fixing the $K^-K^+$ width, and fixing the range of reduced
proper lifetime over which the fits are done. Estimates of the
systematic errors arising from these sources are summarized in
Table~\ref{systerr}. In obtaining systematic errors on $\Delta \Gamma$
from those on the lifetimes, correlation effects are included.  For
example, different production models lead to similar changes in
acceptance for the $K^-K^+$ and $K^-\pi^+$ modes, resulting in a
cancellation in the systematic error in $\Delta \Gamma$. On the other
hand, the uncertainties in mass resolution for the $K^-K^+$ and
$K^-\pi^+$ samples come from fits to independent data samples and are
uncorrelated, and thus their contribution to the uncertainty in $\Delta
\Gamma$ is calculated as the quadratic sum of contributions coming from
the uncertainties in the two lifetimes.  Systematic uncertainties due
to the different absorption cross-sections for $K$ and $\pi$ are
negligible, both because the momentum distributions are quite similar
and because the targets are very thin.

In summary, we measure $\tau_{K\!K}= 0.410 \pm 0.011 \pm 0.006$ ps and
$\tau_{K\!\pi} = 0.413 \pm 0.003 \pm 0.004$ ps. These results are
compared with previous measurements and the world average in
Table~\ref{prevmeas}.  We find $\Delta \Gamma = 0.04 \pm 0.14 \pm 0.05$
ps$^{-1}$, leading to a limit of  $-0.20  < \Delta \Gamma < 0.28$
ps$^{-1}$ at 90\% CL\@. The value of $\Delta \Gamma$ is consistent with
zero and thus, at our level of sensitivity, is consistent with the
Standard Model. This measurement gives us, independent of mixing
studies, a limit on $\Delta \Gamma$.  This 90\% CL limit on the value
of $\Delta \Gamma$ does not saturate the 90\% CL limit on $r_{\rm
mix}$. If $r_{\rm mix}$ were at its limit, this would have to be due to
a contribution from $(\Delta m)^2$.

We gratefully acknowledge the assistance of the staffs of Fermilab and
of all the participating institutions. This research was supported by
the Brazilian Conselho Nacional de Desenvolvimento Cient\'\i fico e
Tecnol\'ogico, CONACyT (Mexico), the Israeli Academy of Sciences and
Humanities, the U.S. Department of Energy, the U.S.-Israel Binational
Science Foundation, and the U.S. National Science Foundation. Fermilab
is operated by the Universities Research Association, Inc., under
contract with the United States Department of Energy.

\bibliographystyle{unsrt}

\newpage

\newpage
\begin{table}[t]
\begin{center}
\caption{Contributions to the systematic errors in lifetimes and lifetime
difference.}
\label{systerr}
\begin{tabular}{|l|cc|cc|cc|}
   \multicolumn{1}{|r}{Systematic error in} & 
   $\tau_{K\!\pi}$ (ps) & &  
   $\tau_{K\!K}$ (ps) & & 
   $\Delta\Gamma$ (ps$^{-1}$) & \\
\hline
Fit Range              & 0.002 & & 0.003 & & 0.024 & \\
Selection Criteria     & 0.001 & & 0.002 & & 0.020 & \\
Particle ID Weighting  & 0.001 & & 0.003 & & 0.024 & \\
MC Production Model\phantom{MMM}
                       & 0.003 & & 0.003 & & 0.000 & \\
Fixed Width            & 0.001 & & 0.002 & & 0.030 & \\
\hline
Total                  & 0.004 & & 0.006 & & 0.050 & \\
\end{tabular}
\end{center}
\end{table}

\begin{table}[hb]
\begin{center}
\caption{Comparison of our measured lifetimes with previous measurements.}
\label{prevmeas}
\begin{tabular}{|l|l@{${}\pm{}$}c@{${}\pm{}$}ll|l@{${}\pm{}$}c@{${}\pm{}$}ll|}
\hspace*{1.25in}&    \multicolumn{3}{c}{$\tau_{K\!\pi}$ (ps)} & &
                  \multicolumn{3}{c}{$\tau_{K\!K}$ (ps)} & \\
\hline
E791            & 0.413 & 0.003 & 0.004 & & 0.410 & 0.011 & 0.006   & \\
E687 \cite{e687}& 0.413 & 0.004 & 0.003 & & \multicolumn{3}{c}{---} & \\
E691 \cite{e691}& 0.422 & 0.008 & 0.010 & & \multicolumn{3}{c}{---} & \\
PDG \cite{pdg}  & 0.415 & \multicolumn{3}{l|}{0.004} &  
                          \multicolumn{3}{c}{---} & \\ 
\end{tabular}
\end{center}
\end{table}

\end{document}